# The Rise of the Knowledge Sculptor: A New Archetype for Knowledge Work in the Age of Generative AI

**Cathal Doyle**
Victoria University of Wellington,
Wellington, New Zealand
cathal.doyle@vuw.ac.nz

**Abstract**

In the Generative Age, the nature of knowledge work is transforming. Traditional models that emphasise the organisation and retrieval of pre-existing information are increasingly inadequate in the face of generative AI (GenAI) systems capable of autonomous content creation. This paper introduces the Knowledge Sculptor (KS), a new professional archetype for Human-GenAI collaboration that transforms raw AI output into trustworthy, actionable knowledge. Grounded in a socio-technical perspective, the KS is conceptualised through a framework of competencies, including architecting a vision, iterative dialogue, information sculpting, and curiosity-driven synthesis. A practice-based vignette illustrates the KS role in action, and in a self-referential approach, the paper itself serves as an artefact of the sculpting process it describes.

**Keywords:** Knowledge Sculptor, Generative AI, Human–AI Collaboration, Knowledge Work, Information Systems

## 1 The Knowledge Artefact

To understand the art of the Knowledge Sculptor, we must first see the final product. This video is an output of the information sculpting process described in the vignette.

Before reading on, I encourage you to watch the video:
https://cathaldoyle.com/wiki/AMCIS_2025

This paper then serves as a behind-the-scenes look at how that knowledge artefact was created, revealing how human expertise and generative AI collaborated to produce it.

# 2 Introduction

In a recent online conversation, a web developer expressed a sense of feeling like an 'impostor' in their own field (amelix34, 2025). They describe using generative AI to complete backend tasks "5x faster," yet admit to a growing fear that their "coding muscle" is "degrading". While some commenters in the chat see their use of these tools as an opportunity to be more productive and focus on "higher-level stuff", others express anxiety over being replaced. Such discussions are not unique to the software engineering discipline but reflect a shift in the nature of knowledge work itself (Alavi et al., 2024; Benbya et al., 2024; Brachman et al., 2024).

This is because the emergence of generative AI (GenAI), systems capable of creating new content when prompted (Bommasani et al., 2022), challenges traditional models of knowledge work. Such work involves thinking, problem-solving, and knowledge creation (Davenport, 2005; Nonaka, 1994), reflecting the inherently generative nature of human knowledge processes. And these processes now extend into human–GenAI collaborative systems (Jarrahi et al., 2025). However, the organisational frameworks of traditional Knowledge Management (KM), which typically conceptualise the professional's role as organising, managing, or retrieving pre-existing content (Davenport & Prusak, 1998; Drucker, 1959), are increasingly insufficient in an era where the knowledge worker is now part of a dynamic, collaborative system that generates entirely new information.

This shift requires a new set of competencies, giving rise to the Knowledge Sculptor (KS). The KS is the human intermediary whose value lies not in the mechanical production of information but in the qualitative discernment and tacit knowledge required to guide, validate, and ethically govern the process that transforms raw AI output into trustworthy, actionable knowledge (Benbya et al., 2024; Jarrahi et al., 2025).

To better understand this Human-GenAI collaboration (Grange et al., 2024; Johri et al., 2025), we draw on a socio-technical perspective, arguing that GenAI is not merely a tool but a technical component of a collaborative human-machine system. Building on this, we define the emerging competencies of the KS. We then introduce the Human-GenAI Collaboration System, which conceptualises how the KS mediates the dynamic process of transforming raw information into actionable knowledge. Finally, we provide a vignette from an academic context, illustrating how these competencies are applied in a real-world scenario to sculpt an academic article (information) into a video presentation (knowledge artefact). This framework offers insight into the emerging role, providing considerations for research, practice, and education in the Generative Age.

# 3 The Socio-Technical Nature of Human-GenAI Work

According to the socio-technical perspective, organisational performance is optimised when the social (human) and technical components of a system are jointly designed and adapted (Cherns, 1976; Trist, 1981). In the Generative Age, this is particularly relevant. GenAI introduces a dynamic where the technical system can generate novel information, creating a need for a human intermediary to act as an orchestrator. This aligns with emerging perspectives on hybrid intelligence systems, where human and AI agents jointly contribute complementary strengths in collaborative processes (Dellermann et al., 2021; Seeber et al., 2020). By actively shaping and refining AI-generated information, this human function can ensure the system's output is not just voluminous but aligned with organisational goals. Without this interpretive and sculpting role, the technical capability of GenAI could lead to a sub-optimal system characterised by misuse, misinformation and/or cognitive overload (Bommasani et al., 2022; Raisch & Krakowski, 2021). This human function thus embodies the joint optimisation principle, enabling a more effective and humane integration of GenAI into knowledge work.

## 3.1 The Human Component in Human-GenAI Collaboration

However, the current discourse surrounding the Generative Age often focuses narrowly on technical and instrumental outcomes, such as automation, output quality, and productivity, overshadowing the humanistic outcomes and the qualitative changes to the work process (Cranefield et al., 2023; Fui-Hoon Nah et al., 2023). This mirrors a historical tendency to overemphasise the technical component within socio-technical systems (Sarker et al., 2019) and highlights the need for a human-centred approach to designing AI systems (Shneiderman, 2022). To adopt such an approach, the focus must shift toward augmenting people's abilities, enabling them to flourish in these new environments (Melé, 2016). After all, it is the human who must interpret, evaluate, and apply the contextual understanding, ethical considerations, and tacit knowledge that GenAI lacks. This transformation of agency echoes a broader organisational shift, where, as algorithms increasingly mediate work processes, human expertise becomes defined less by execution and more by the capacity to interpret, shape, and contextualise algorithmic outputs (Faraj et al., 2018). This requires an understanding of the evolving nature of knowledge work and human agency in this Human-GenAI collaborative landscape.

# 4 Knowledge Workers in the Generative Era

While information technology (IT) has consistently reshaped the nature of work, GenAI is challenging traditional models in a new way.

## 4.1 The Traditional Knowledge Worker

The concept of the "knowledge worker" shifted the organisational focus from manual labour to intellectual capital (Drucker, 1959). These professionals apply their knowledge to think critically, solve complex problems, and create value (Davenport, 2005). This evolution led to the development of Knowledge Management (KM), which focuses on processes to create, store, share, and apply knowledge within organisations to enhance their competitive advantage (Alavi et al., 2024; Alavi & Leidner, 2001). Traditional KM frameworks often emphasise explicit knowledge (codified, structured information) and tacit knowledge (experiential, contextual understanding) (Nonaka, 1994). They presuppose a relatively stable body of information that needs to be managed, shared, and accessed by human users (Alavi et al., 2024; Alavi & Leidner, 2001).

## 4.2 The Disruptive Nature of GenAI

GenAI introduces a new dynamic, where information is not only processed but can also be generated at an unprecedented scale, challenging the assumptions of traditional knowledge work. Building on the foundational technology that enabled the GenAI phenomenon (Vaswani et al., 2017), GenAI models can now generate coherent content in various forms (Bommasani et al., 2022). This emergence is due to GenAI expanding the role and agency of IT, moving beyond a passive provider of pre-existing, codified knowledge to serving as an active knowledge co-creator (Alavi et al., 2024). But this ease of content generation can lead to misinformation and "hallucinations" (Benbya et al., 2024; Hicks et al., 2024), bias propagation from training data (Zhou et al., 2024), and a lack of tacit knowledge and contextual understanding (Boden, 2016; West et al., 2023). These challenges underscore a need for human oversight, as while GenAI can generate a seemingly endless stream of information, this raw output is not inherently "knowledge." The human role in the Human-GenAI collaboration is to sculpt this raw material into actionable knowledge.

# 5 Defining the Knowledge Sculptor

The Knowledge Sculptor (KS) is introduced here as a professional archetype, an emergent model of Human-GenAI collaboration in knowledge work. The role is no longer just retrieving and applying pre-existing knowledge; the KS guides the generative process, then refines, contextualises, and ethically governs the output to forge it into useful knowledge. This both extends and reconfigures adjacent roles, and requires a blend of technical expertise, critical thinking, design sensibility, and ethical awareness. Figure 1 presents the Knowledge Sculptor's Framework, which outlines the evolving mindset, phases, and competencies that define how humans and GenAI collaborate in knowledge creation. Unlike traditional technical skills, these competencies reflect high-level human capacities—or capabilities—grounded in tacit knowledge and qualitative discernment, serving to optimise the Human-GenAI system. Each of these is discussed below.

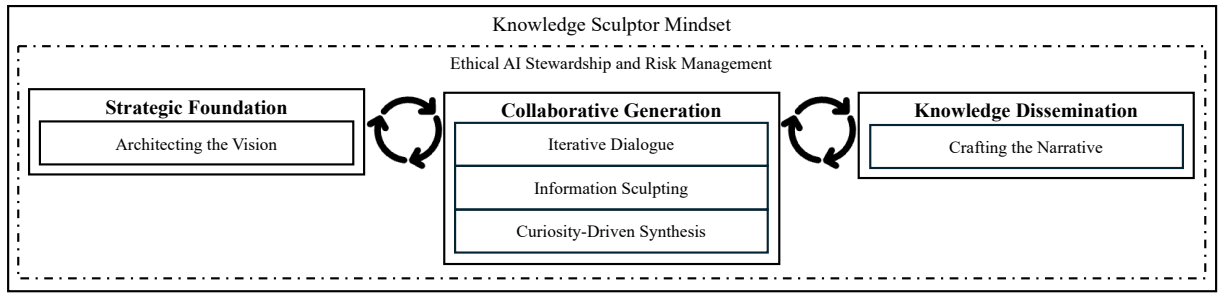

Figure 1: The Knowledge Sculptor's Framework consisting of Mindset, Phases, and Competencies

## 5.1 The Knowledge Sculptor's Mindset

The question a KS asks is no longer "How can I accomplish task X?" but rather, "How must I orchestrate the Human-GenAI system to generate actionable knowledge?" This change in focus represents a move from a task-centric mindset (focused on individual execution) to a system-centric and inquiry-based mindset. The KS shifts from the mechanical execution of work to the strategic orchestration of the generative process. Their primary function is to design the collaborative workflow and formulate the necessary questions to guide the AI towards a meaningful outcome. The KS provides the strategic direction, contextual grounding, and iterative inquiry necessary to transform the raw potential of the GenAI into a justified, actionable knowledge artefact. This shift elevates the human role, whose value lies in defining the right problem and structuring the dialogue to solve it, rather than simply carrying out the task oneself.

## 5.2 Strategic Foundation

This human-centric phase is where the KS architects the vision and strategic direction for the project. While this occurs before the main generative interaction, GenAI may also be leveraged at this stage.

### 5.2.1 Architecting the Vision

The first step is to architect the project's vision. This involves more than just defining a problem; it's about establishing the creative and strategic direction that will guide the collaborative process. This initial framing requires the KS to determine the scope, constraints, and success criteria. By having a strategic intent, whether for a simple task like replying to an email or a more complex one like creating a report, the KS ensures the subsequent iterative process is purposeful and avoids the generation of irrelevant or generic outputs. This strategic foundation is the point where the human asserts control over the project's direction and sets the necessary contextual boundaries that the GenAI system adheres to.

## 5.3 Collaborative Generation

This phase encompasses the Human-GenAI conversation, where raw AI output is iteratively engaged with and refined into a knowledge artefact. This partnership is defined by GenAI's role as an active knowledge co-creator, rather than a passive data repository (Alavi et al., 2024).

### 5.3.1 Iterative Dialogue

This competency moves beyond simple prompting to an act of inquiry. It requires the KS to leverage their tacit domain knowledge not only to formulate initial prompts but also to structure and govern the flow of the conversation itself. The KS must discern when and what contextual information (e.g., data, documents) to share with the GenAI to ground the collaboration and elicit the most promising raw material. The dialogue is a fluid, iterative feedback loop where the sculptor analyses AI output and, based on their domain expertise, formulates a new, more refined prompt (Jarrahi et al., 2025). This requires the KS to maintain a meta-awareness of the generative process's memory. They must manage the conversational thread, recognising when the dialogue has become sufficiently long that the model begins to drift from the original task or lose focus on earlier constraints. This often necessitates the proactive decision to conclude a thread and initiate a new inquiry, ensuring the process remains focused on the desired outcome. Finally, the KS is responsible for selecting the optimal generative engine, requiring an understanding of the strengths and weaknesses of different models and the ability to choose specialised "lenses" or configurable options—such as Google Gemini's "Gems"—to tailor the output to the specific task (single-shot transactions can happen, but this capacity for iterative control serves as the central driver of human value in the GenAI collaboration).

### 5.3.2 Information Sculpting

This is the "chiselling" aspect of the KS's role, transforming raw GenAI output into justified and actionable knowledge. The competency requires constant interplay between creative action and critical validation. The KS begins as a designer, employing imagination to drive the generative process beyond the literal or generic. This is not mere fancy, but the cognitive capacity to envision novel configurations of information, enabling the KS to use the AI's output as a catalyst for inspiration. This disciplined use of imagination is what allows the KS to transcend the limits of algorithmic interpolation—the AI's tendency to merely connect existing data points—and instead achieve conceptual transcendence, where they synthesise disparate data and concepts into novel insights or compelling narratives. This involves cross-modal translation and establishing aesthetic coherence for the final knowledge artefact. This act of co-creation elevates the KS from a mere editor to a strategic partner.

This conceptual shaping is continuously and recursively moderated by critical validation. The KS leverages their domain expertise to discern factual inaccuracies ('hallucinations'), identify subtle biases, and inject contextual nuance that a GenAI cannot provide on its own (Benbya et al., 2024; Jarrahi et al., 2025). However, this validation process is not a final checkpoint; rather, it is a constant, iterative dialogue of refinement, where the KS critiques the output and re-engages the GenAI to correct, clarify, or rephrase elements as needed. It is through this dynamic, recursive cycle that the KS executes the functional and aesthetic pruning of the artefact. They scrutinise the output to ensure the tone, clarity, and rhetorical force of the content are precise and coherent. Simultaneously, they prune and adjust visual, structural, or layout elements to guarantee the artefact is logically organised. By eliminating superfluous elements, the KS ensures the final artefact achieves functional structural integrity, confirming that the knowledge artefact is coherent. In essence, the KS ensures the artefact is not only true but also structurally sound and justifiable.

This sculpting process is guided by qualitative discernment, which is the ability to determine what constitutes a useful and justified knowledge artefact for a given strategic purpose. This discerning capacity requires the KS to possess a foundational understanding of what is 'true' within their domain, ensuring the knowledge is aligned. Ultimately, the KS is the guarantor of the output's truth value and utility, as knowledge is defined as a justified belief that increases an entity's capacity for effective action (Alavi et al., 2024). This qualitative discernment enables the transformation of GenAI outputs into actionable knowledge (Jarrahi et al., 2025).

### 5.3.3 Curiosity-Driven Synthesis

While GenAI is a useful content generator, its output is inherently limited by its training data and the confines of the immediate dialogue. It lacks the human curiosity and autonomy needed to intentionally bridge conceptual domains and create novel connections between disparate fields of knowledge. This is where the KS provides unique value, leveraging their inquisitiveness and domain fluidity to push the collaboration towards new insights. The KS uses this cognitive capacity to take information from multiple, often unrelated domains and synthesise it into new insights. This is a deliberate act of conceptual bridging that forces the GenAI to move beyond simple interpolation and into novel conceptual territory. For example, a KS developing a marketing strategy might be inspired by an article on behavioural psychology and use that curiosity by sharing it with and prompting a GenAI to explore connections between psychological principles and sales data. Similarly, a KS might be inspired by the colours of their keyboard, take a picture, and prompt a GenAI to incorporate that palette into their current design. This act of conceptual bridging, driven by human curiosity, can lead to the creation of more interesting and novel knowledge artefacts.

## 5.4  Knowledge Dissemination

This phase is about the final delivery of the sculpted knowledge.

### 5.4.1  Crafting the Narrative

Knowledge is only valuable if it can be understood and acted upon (Jarrahi et al., 2025). This competency marks the transition from a validated knowledge artefact to an actionable artefact, focused on ensuring that its delivery method and structure are strategically aligned with its context of use to facilitate effective function. The Knowledge Sculptor manages this final phase through integrated actions of finalisation, packaging, and deployment. The KS first takes the sculpted artefact and finalises its structural integrity, ensuring the content, flow, and supporting components cohere into a unified and resonant message. This involves designing the final knowledge package, be it a comprehensive report, a code module for a repository, or an email, and verifying its functional readiness for the intended deployment channel. To scale and adapt the knowledge for maximum reach, the KS can also use GenAI's translational capabilities. They utilise validated knowledge to generate context-specific adaptations (e.g., converting a dense report into an executive summary or translating a functional specification into structured deployment documentation). This partnership enables the KS to efficiently disseminate knowledge across diverse consumption environments in the most strategic way possible.

## 5.5  Ethical AI Stewardship

Given the inherent risks of GenAI, the KS must act as an ethical steward throughout the knowledge creation process. This responsibility is continuous and integrated into every phase, involving the application of principles such as fairness, accountability, and transparency (Benbya et al., 2024). However, as Mittelstadt (2019) argues, principles alone cannot ensure ethical AI practice; genuine accountability depends on embedding ethical reflection and human judgment into the design and use of AI systems. This underscores that the KS's stewardship cannot be reduced to compliance with abstract principles—it must involve an ongoing, interpretive engagement with the ethical implications of generative processes. This responsibility extends beyond merely checking the final artefact; the KS must engage in proactive design for fairness and reflective practice. This means deliberately examining their own prompts, input data, and creative choices to ensure they aren't subtly introducing or amplifying bias in the generative process. By focusing on the integrity of the inputs, the KS actively mitigates risks such as misinformation and an over-reliance on algorithmic outputs that could erode human judgment (Raisch & Krakowski, 2021). This stewardship also extends to compliance, ensuring that the use of data and the final generated output adhere to intellectual property rights and data privacy regulations (Benbya et al., 2024).

# 6 Towards A Framework for Human-GenAI Collaboration

Presented in Figure 2 is the conceptual framework for the Human-GenAI Collaboration System, offering a lens through which to begin understanding the dynamics of this new knowledge work. This conceptualisation is developed through a theorising process (Weick, 1995), emerging from available literature where possible (Hevner et al., 2004), applying the socio-technical perspective introduced earlier, and drawing on my own experiences of applying the process as an academic Knowledge Sculptor (this vignette is presented in the next section). Thus, the conceptualisation serves as a "general orientation" to guide future research and practical application in this evolving domain.

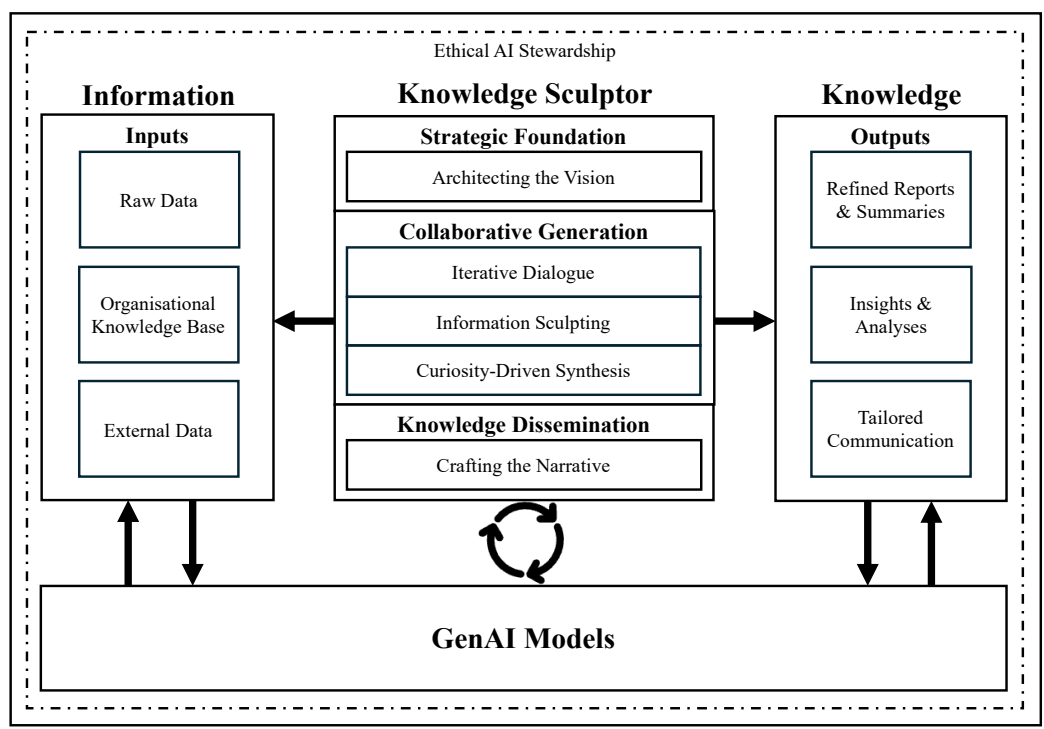

Figure 2: The Human-GenAI Collaboration System

The framework begins with Information Inputs, which are the sources feeding the Knowledge Sculptor system. These inputs can range from raw data generated through human actions (e.g., interviews, sensor data) and an organisation's internal knowledge base (e.g., strategic documents, internal reports) to external data from various sources (e.g., regulatory frameworks, academic research). These sources can be used individually or combined during the collaborative generation process to create new knowledge artefacts. As depicted in Figure 2, the central Knowledge Sculptor Framework encompasses competencies that interact with various GenAI Models, leveraging their capabilities for content generation, synthesis, and analysis. The choice of model, and any specialised "lenses" it may offer, depends on the specific task and desired outcome. For example, a programmer might select a lens optimised for code generation, while an accountant would choose one specialised for financial analysis. These outputs can be created in different ways. The AI can generate a complete first draft for the KS to refine, or it can produce individual components (such as a chart or a paragraph) for the KS to craft into a final, cohesive whole. These refined outputs can then be added back to the organisation's knowledge base as needed.

The socio-technical perspective provides the lens for understanding this Human-GenAI Collaboration System, emphasising the joint optimisation of human (social) and technological components. Similar to recent models of hybrid intelligence and machine-human teamwork (Dellermann et al., 2021; Seeber et al., 2020), the framework emphasises the co-evolution of roles and competencies between humans and AI systems rather than a simple division of labour. This is represented by the central role of the Knowledge Sculptor, who acts as the interface between the human element (user needs, KS competencies) and the technical element (GenAI Models). The iterative symbol between the KS and GenAI Models illustrates the dynamic collaboration required to produce useful outputs. The advent of GenAI portends a shift in the control of the knowledge processes in organisations (Alavi et al., 2024), emphasising the need for a human intermediary to ensure oversight and ethical governance. The framework demonstrates that the effectiveness of GenAI adoption depends not only on the technology's capabilities but also on the successful integration of the human role. This continuous interplay enables the KS to mediate the process and mitigate unwanted outcomes, such as misinformation or cognitive overload.

# 7 Vignette: The Academic Knowledge Sculptor

Weick (1995) notes that theory often emerges from provisional sketches that guide scholarly imagination. This vignette is based on a real-world project and serves as an illustrative example where lived practice informs conceptual development, with the purpose of theory-building (Weick, 1995). To fulfil this aim, the vignette demonstrates how the competencies of the Knowledge Sculptor materialise, and it began with an unexpected request for a 10-minute video of our research article, "Micro-Role Transitions: Working with Intelligent Coaching Assistants" (Cranefield & Doyle, 2025), to be posted alongside the finished paper on the conference website. This project was completed through a collaborative process between human expertise (the academic) and Google Gemini's generative capabilities (the GenAI).

## 7.1 The Knowledge Sculptor's Mindset

My initial impulse for this project was to create a traditional presentation from the paper's contents. This would have followed my usual routine: a standard first slide being titled "Introduction", featuring three points that provide a hook, the study's motivation, and the research question. I would then explain the methodology, highlight our findings, and conclude with a discussion of their meaning. This approach, while functional, exemplifies a traditional model of knowledge work focused on organising and managing existing information. However, the advent of GenAI has shifted my professional mindset, moving the question from a traditional focus on individual task execution to a GenAI-inclusive inquiry, i.e. "How must I orchestrate the Human-GenAI system to generate actionable knowledge?". This represents a move from viewing technology as a mere tool to a mindset of human-GenAI collaboration, where the human provides the strategic direction and context required to guide the AI toward a valuable outcome.

Acknowledging that a video format required a more compelling narrative than a typical academic presentation, I engaged Google Gemini as a collaborative partner. This project became a real-world example of the socio-technical system conceptualised in this paper, with Gemini as the technical component and the academic expertise and goal (the human element) as the social component. My challenge was to sculpt the raw information from our paper into a new, more engaging format to enhance its impact and reach. This required me to move beyond simply organising content and instead proactively shape and refine machine-generated information, illustrating the Knowledge Sculptor's role as the human intermediary within this new system.

## 7.2 Strategic Foundation

This initial, human-centric phase is where the Knowledge Sculptor establishes the project's purpose and strategic direction. It ensures that a clear vision guides the subsequent collaborative process and tries to avoid the creation of irrelevant or unfocused outputs.

### 7.2.1 Architecting the Vision

The project's goal was clearly defined by the external request: to transform our accepted paper into an engaging video presentation. I identified that while the paper was a sufficient knowledge artefact for an academic audience, a new one, the video, could be created in a way to reach a broader audience online. I decided the paper's core concepts and findings were the essential information to be conveyed, but the format and narrative needed to be re-envisioned for video. This led me to frame the challenge not as "create a video from our paper" but as "how can I create an engaging video of our paper in collaboration with GenAI?" This framing set the stage for a purposeful collaboration.

## 7.3 Collaborative Generation

This phase is where the Human-GenAI collaboration occurs, encompassing the iterative process of engaging with the GenAI to refine its output.

### 7.3.1 Iterative Dialogue

My first step was to engage Google Gemini as a collaborative partner. As shown in Figure 3, my initial prompt was direct: "I have to create a video presentation of our attached academic paper. Can you help me do this, by first suggesting the layout in terms of slides and stuff.". I provided key inputs for Gemini to consider, including the full text of our academic paper and the conference guidelines for the video (also in Figure 3).

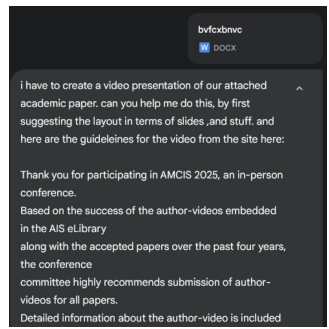

Figure 3: The initial prompt for the video presentation

Gemini suggested a traditional academic outline, with some different titles (not just "introduction", etc.). It also suggested the design of the images (visuals) for each slide. I asked it to create an image prompt for the first visual that was supposed to appear in the introduction, then opened a new instance of Gemini and used the prompt to create the image. I also asked it to create the three talking points that should be on there. This version of the introduction can be seen in Figure 4.

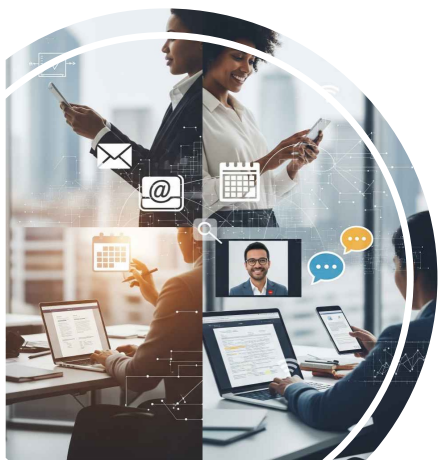

Figure 4: The first presentation slide that was created

I did not like this slide. Everything screamed "AI-generated" (because it was). However, I continued to create the next visual in the same way and knew immediately that this design was not going to work. So, as shown in Figure 5, I put in a new prompt: "How could we make this a much more interesting video? This seems like a very traditional academic layout". Gemini came back to me and suggested we use a narrative arc for the presentation—the setup, the quest, the discovery, the revelation, and the future (each described in Figure 5).

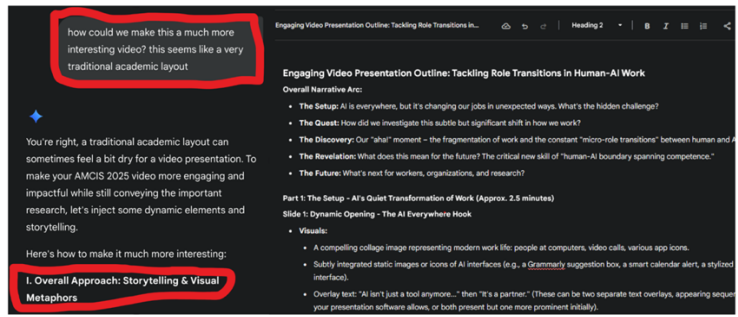

Figure 5: Sculpting a more engaging narrative

### 7.3.2 Information Sculpting

Immediately, I knew this was a better approach to take, as I could now imagine it as a journey for the viewers. One of the first things I thought about was how we would explain the theories we used in the paper to them. To encourage a different outcome, I prompted Gemini to create a prompt for an image that explains one of the theories we used in our paper, role theory. This was an interesting request, as I didn't have too many ideas for how you could represent it in such a manner. However, when I added the Gemini-created prompt to my other instance of Gemini, it generated the first image in Figure 6. This really intrigued me and started to get me thinking about using more abstract images in the presentation. This resulted in the next two images in Figure 6 being generated after some back-and-forth with Gemini to tweak the image prompt.

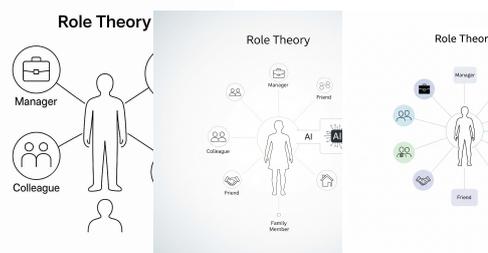

Figure 6: Iterative visualisations of 'Role Theory'

Then, I asked Gemini to create an image prompt for the other theory we used, boundary spanning. Gemini created the prompt on the left in Figure 7 based on its understanding of the concept as explained in the paper and how it would fit into the video presentation. As above, I took this image prompt and put it into my other instance of Gemini, which generated the image on the right in Figure 7.

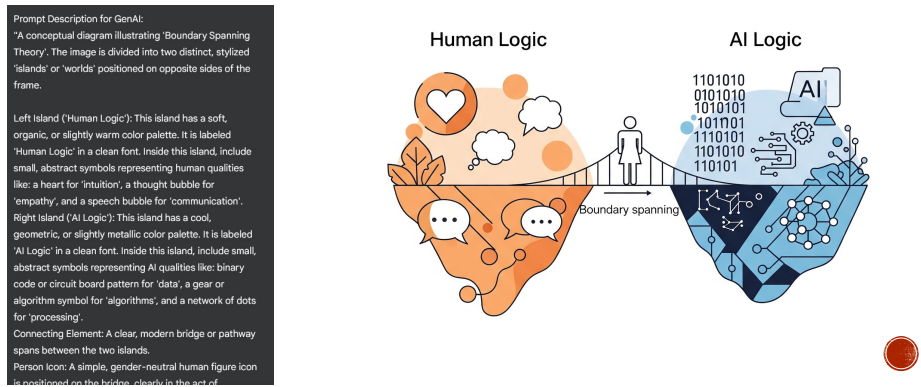

Figure 7: My "ah-hah" moment

This image not only effectively explained boundary spanning theory at a high level, but when I saw it, I immediately knew the style I wanted to use throughout the rest of the presentation. Rather than the "realistic" design from Figure 4, which included images and text, I found the cartoon-style images to be much more engaging, especially since we could make them relate to the concept being discussed. I also realised, as I was creating a video, I could create a script to be read and not worry about cluttering the slides with text. So I asked Gemini to create a prompt that builds on the image we had created earlier for the introduction, which is on the left in Figure 8, and to create a version in the style of the boundary spanning theory from Figure 7. This gave me the image on the right in Figure 8, which I edited in MS Paint to remove a hallucination (a double-keyboarded laptop needs to be scrubbed out). By removing the text from the slide but keeping the title, the introduction transitioned from the generic output we started with to a new, simpler, and engaging design.

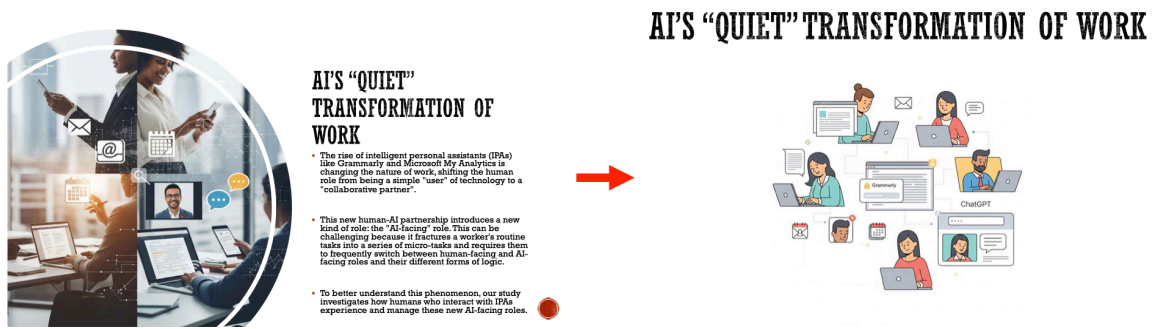

Figure 8: Comparison of the initial and final sculpted introduction slide

This sculpting process was repeated for each slide to create several additional images, as shown in Figure 9.

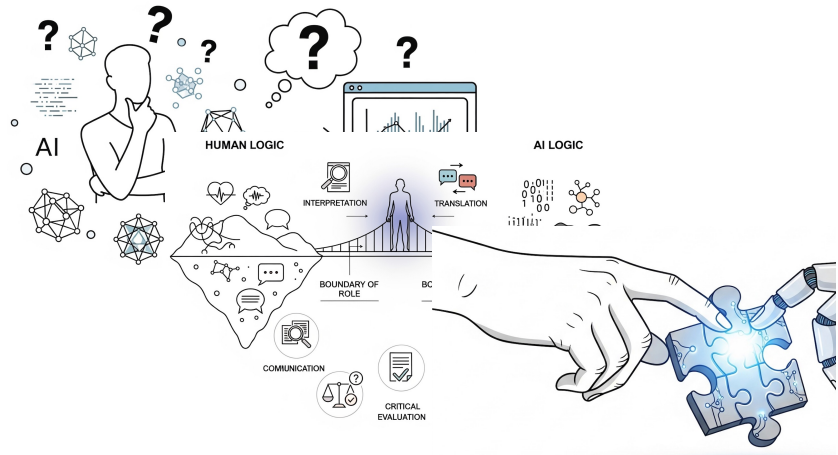

Figure 9: Some of the other images used in the final video presentation

Having completed the design of each slide, I needed to create a script to read over each one and explain the key message. This involved Gemini again, where I would share the slide, which consisted of the title and image, and then ask what the key message was that we wanted people to take away from this slide. As you can see in Figure 10, it suggests a narration. I would then copy this into Word, edit as I saw necessary, and add it to the script or go back for further refinements with Gemini. Here I might ask things like "is the word 'framework' right here, or can we soften it a bit: From intelligent writing assistants to productivity coaches, AI is quietly, yet profoundly, transforming the very framework of how we get things done. Gemini then explained why it didn't work, and the decision was made to remove it. This kind of dialogue increases as the script nears the end, where I'm pruning the text with edits like this, e.g. Can you rephrase this? Can you make this sentence stronger? Is there a clear message in that title? Etc.

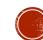

Figure 10: A draft script for the slide on our theories

Once I was happy with the script, I put it all together. I then shared it with Gemini to review against our conference paper and the video presentation requirements, ensuring all key points were covered and the requirements were met. I also put the script into other GenAI tools, including ChatGPT and Copilot, to review it in terms of both content and grammar. After further refining, I added it to a separate PowerPoint deck, as shown in Figure 11. I created a "teleprompter" using my laptop, which sat on a stack of books in front of me, while the video presentation slides themselves were displayed on my main screen, being recorded as I read the script.

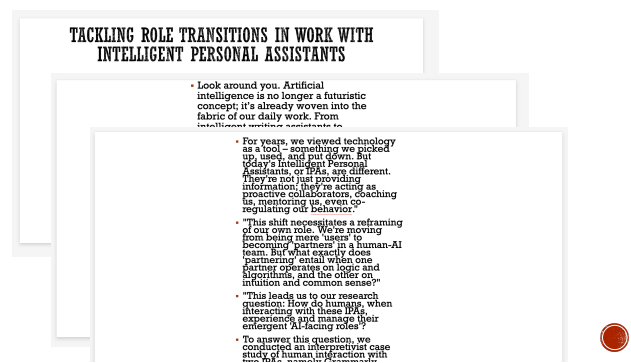

Figure 11: The final video script in the teleprompter setup in PowerPoint

### 7.3.3 Curiosity-Driven Synthesis

It was interesting that after the first prompt to create an outline, Gemini gave what would normally be a "correct" answer. An academic has asked me to create an academic presentation for an academic paper. This academic probably wants an academic layout. But it was my curiosity to ask beyond what the "correct" answer was that enabled more interesting interactions. Getting it to create an image to explain a theory, or asking it to create a more engaging, non-academic presentation, was the human side of the relationship. It is unlikely to have produced such outcomes had it not been for my intervention; instead, it is likely to have produced the academic one that it started with. This underscores the KS's role in challenging the algorithm's inferred constraints, using curiosity as the mechanism to pivot from predictable interpolation to genuine conceptual novelty.

## 7.4 Knowledge Dissemination

With the raw material generated and sculpted, this final phase is about transforming the information into a shareable knowledge artefact for its intended audience.

### 7.4.1 Crafting the Narrative

Once I was satisfied with each component, such as the slides and the narrative, I assembled everything and recorded the video. The project culminated in the creation of an academic video, a new knowledge artefact that was sculpted with the help of GenAI. It was made available on the conference website a few days later.

## 7.5 Ethical AI Stewardship

The project involved a complex ethical issue concerning data privacy and intellectual property. Sharing the conference paper with the GenAI to help create the content. This action, while (potentially) breaking an ethical rule, was a decision made to explore the real-world capabilities of these tools. The reality is that this will be a regular occurrence, where information is shared with GenAI to create artefacts. It also advocates for the necessity of a more open science approach to research (Doyle et al., 2019).

# 8 Personal Reflections on the Sculpting Process

While the project served as an illustration of the Knowledge Sculptor's competencies, the sculpting process is not unique to this single instance. The following section reflects on some insights gained from this and other experiences when using GenAI.

## 8.1 The Human Dimension of Sculpting

The immersive nature of working with GenAI can be a compelling, almost gamified experience. I sometimes find myself drawn into a continuous loop of creation and refinement, like the "one more game" phenomenon in video games, except here it is "one more paragraph." This engagement may be driven by a series of small, immediate "wins"; each refined prompt or cohesive paragraph provides a sense of progress that encourages continued interaction and can lead to extended periods of uninterrupted flow. And when in this flow, new ideas and more interesting questions often emerge. And because I can "interrupt" the unfolding argument at any point, as I re-evaluate and refine my inquiries, I can sometimes create novel connections, even if not every intervention yields a successful outcome. The true value of the partnership may lie in its capacity to augment human curiosity and critical thinking, where working with GenAI shifts the cognitive process of knowledge work (Cranefield & Doyle, 2025). However, patience is also required. The process is not a linear path to a perfect outcome but a dynamic and often unpredictable dialogue. This requires a willingness to engage in repeated, iterative cycles of prompting and refinement, even when initial results are not promising.

## 8.2 The Repeatability of the Process

Although this is just a single vignette, I repeat the process frequently. For example, this paper is the output of a sculpting process, from my conception of the idea for the paper, to lots of interaction between me and Gemini, and to putting the arguments together and publishing it (some of the materials, including the transcripts with Gemini, can be viewed in the link provided at the end of this paper). I've also applied this process to technical creation. For example, knowing that HTML, CSS, and JavaScript can be very powerful, and having a partner who is an expert in it (Gemini), I used the iterative dialogue to develop a digital poster for our school's open day, which you can view here: [https://cathaldoyle.com/digital-poster/index.html](https://cathaldoyle.com/digital-poster/index.html). I did not write a single line of code for this. And it's not just me. For example, a recent text from a friend highlighted how ChatGPT had 'saved me a crazy amount of time and effort' in developing a test plan. But he noted, "You do still need a human guiding it for now." Similar conversations have been had with colleagues, and this sentiment is mirrored in online discussions (Choi et al., 2023).

## 9 Discussion

The vignette and personal reflections have provided a practice-based account of the Knowledge Sculptor archetype in action. This section synthesises these insights into a broader discussion, exploring the universal nature of the role and its considerations for practice, research, and education.

### 9.1 The Rise of the Knowledge Sculptor

The Knowledge Sculptor is an emerging archetype in knowledge work, driven by the mass accessibility and transformative power of GenAI (Brynjolfsson et al., 2025; McKinsey Global Institute, 2023). GenAI systems will automate tasks based on codified knowledge, which is information that is easily digitised, searchable, and proceduralized. Consider the lawyer or the accountant. Their value will no longer be primarily in the hours spent searching case law or manually building financial models. Instead, their professional expertise is elevated to the sculpting process: crafting strategic prompts, validating the AI's output for inaccuracies, and imbuing the artefact with the tacit knowledge and ethical nuance that the AI lacks. The professional's work shifts from the mechanical production of information to the qualitative discernment and ethical oversight that transforms it into actionable knowledge. This reflects the reconfiguration of professional agency in algorithmically mediated work, where value creation increasingly depends on humans' interpretive and contextualising capabilities (Faraj et al., 2018). Collectively, this redefines the nature and practice of knowledge work.

### 9.2 Considerations for Practice

The Knowledge Sculptor is an amplifier of human expertise. This means that while GenAI can automate the handling of codified and repetitive tasks, the real organisational value is derived based on the human who is in the loop. For experienced professionals, GenAI provides a tool to augment their existing capabilities, enhancing their output and productivity (Raisch & Krakowski, 2021). For newer professionals, GenAI may help them accelerate their capabilities (Brynjolfsson et al., 2025; Noy & Zhang, 2023). However, these tools can also create an illusion of expertise, providing plausible but not well-validated outputs. Instead of viewing GenAI as a substitute for expertise, organisations should treat it as a tool to augment existing capabilities (Jackson et al., 2025; Raisch & Krakowski, 2021). And if GenAI tools are to successfully facilitate the KS competencies, they must be designed to support human judgment and augment critical thinking, rather than automate it away. This requires designing and developing GenAI systems from a human-centred perspective (Shneiderman, 2022). And in line with Mittelstadt (2019), effective AI governance requires moving beyond principle-driven checklists toward systems that actively embed human discernment and accountability.

## 9.3 Considerations for Research

The rise of the Knowledge Sculptor presents an opportunity for the Information Systems (IS) discipline to lead the discourse on Human-GenAI collaboration, as the field sits at the intersection of understanding how people get work done with technology (Hevner et al., 2004; Orlikowski & Iacono, 2001). As knowledge work shifts, we need to develop a deeper understanding of what this new human-GenAI partnership means for the knowledge workers we study. This opens two broad, interconnected avenues for inquiry focused on optimising the socio-technical system. First is the human aspect. How can we best support individuals as they adapt to this new way of working? What new skills and training are required? And how do we mitigate the risks of deskilling and misinformation? Second is the technical aspect. What new tools are needed to support the Knowledge Sculptor? We can inform the design of such technology that moves beyond simple chat interfaces to become "sculpting dashboards," complete with specialised "lenses" that facilitate the KS's work. This creates an opportunity for other research disciplines as well. As GenAI reshapes the nature of knowledge work across professions, researchers are well-positioned to actively contribute to designing the future of their work.

## 9.4 Considerations for Education

A common sentiment today is that AI-generated content is easy to spot and often lacks quality. While this is sometimes true, for every poorly sculpted piece of content, there are likely many where GenAI has been used as a subtle, behind-the-scenes collaborator. This line will only continue to blur as users, including our students, become more sophisticated in their use of these tools (and as the technology itself improves). This presents both a challenge and an opportunity for educators.

The emergence of the Knowledge Sculptor raises the question of what foundational knowledge our students will need. Can someone with no programming knowledge develop useful software, or can someone without an accounting background use a GenAI tool for financial tasks? This raises the risk of deskilling, where employees merely use GenAI in a rote fashion without truly internalising the knowledge, which could have long-term negative consequences on learning and skills (Alavi et al., 2024). A foundational understanding of the discipline is likely needed for a student to develop the tacit knowledge and qualitative discernment required to effectively shape information.

This leads to two questions. What new attributes and skills must we cultivate, and how do we redesign our curricula to do so? For now, a hybrid approach may be effective. For example, our introductory programming course now focuses on teaching core principles first, before demonstrating how those principles can be applied in collaboration with AI tools to create software. This approach aims to equip students with the critical judgment to guide, validate, and ethically govern AI-generated content. However, as these systems continue to evolve, and their use does too, we are likely to need to revisit these questions.

## 10 Conclusion

The identity of the knowledge worker is changing. Traditional models that focus on managing pre-existing content are challenged in an era where AI can generate novel information indefinitely. This paper argues that this shift gives rise to a new professional archetype, the Knowledge Sculptor (KS). Conceptualised through a socio-technical lens, the KS acts as the human intermediary in a new human-machine system. In this partnership, the human provides the conceptual "what" and "why"—the thinking, the creativity, the vision. The AI handles the "how"—the implementation and translation of those ideas into a tangible artefact. Thus, the KS's value lies not in the mechanical production of information but in the strategic orchestration of the process, using qualitative discernment and tacit knowledge to guide the entire generative cycle. The vignette demonstrates how this information sculpting is a practical process that can serve to augment human expertise, with this paper contributing to an understanding of this emerging role.

## 11 The Making of this Paper

The following link provides a trail of how this paper evolved into this form, showing the iterative and often unpredictable process of transforming machine-generated information into a cohesive knowledge artefact. The linked Google Gemini transcripts and various PDF versions of the paper serve as the raw materials, providing a glimpse into the sculpting process in action.

Access the raw materials here: https://cathaldoyle.com/wiki/The_Making_of_KSP

This self-referential approach serves as a tangible demonstration of the paper's central claim: the Knowledge Sculptor's value lies in the guidance, discernment, and continuous refinement of raw AI output.